# Night sleep duration trajectories and associated factors among preschool children from the EDEN cohort

Authors:

Sabine Plancoulaine, MD, PhD[a,b], Eve Reynaud[a,b,c], MPH, Anne Forhan[a,b], Sandrine Lioret, PhD[a,b], Barbara Heude, PhD[a,b] and Marie-Aline Charles MD, PhD[a,b] ; on behalf of the EDEN mother-child cohort study group.

**Affiliations:**

[a] INSERM, UMR1153, Epidemiology and Statistics Sorbonne Paris Cité Research Center (CRESS), early ORigins of Child Health And Development Team (ORCHAD), Villejuif, F-94807 France;

[b] Univ Paris-Descartes, UMRS 1153, Paris, France;

[c] Ecole des Hautes Etudes en Santé Publique (EHESP), Rennes, F-35043, France

**Address correspondence to:**

Sabine Plancoulaine, INSERM, UMR1153, Epidemiology and Statistics Sorbonne Paris Cité Research Center (CRESS), early ORigins of Child Health And Development Team (ORCHAD), 16 Av Paul Vaillant Couturier, 94 807 Villejuif Cedex, FRANCE,

Email : sabine.plancoulaine@inserm.fr

Phone: + 33 1 45 59 51 09.

**Ethical approval and consent to participate:**

The study was approved by the ethics research committee of Bicêtre Hospital (Comité Consultatif de Protection des Personnes dans la Recherche Biomédicale) and by the Data Protection Authority (Commission Nationale de l'Informatique et des Libertés).

**Funding:**

This research did not receive any specific grant from funding agencies in the public, commercial, or not-for-profit sectors.

**ABSTRACT**

**Objective.** Sleep duration may vary inter-individually and intra-individually over time. We aimed at both identifying night-sleep duration (NSD) trajectories among preschoolers and studying associated factors.

**Methods.** NSD were collected within the French birth-cohort study EDEN at ages 2, 3 and 5-6 years through parental questionnaires, and were used to model NSD trajectories among 1205 children. Familial socioeconomic factors, maternal sociodemographic, health and lifestyle characteristics as well as child health, lifestyle, and sleep characteristics at birth and/or at age 2 years were investigated in association with NSD using multinomial logistic regressions.

**Results.** Five distinct NSD trajectories were identified: short (SS, <10h, 4.9%), medium-low (MLS, <11h, 47.8%), medium-high (MHS, ≈11h30, 37.2%), long (LS, ≥11h30, 4.5%) and changing (CS, i.e. ≥11h30 then <11h, 5.6%) NSD trajectories. Multivariable analyses showed in particular that, compared to the MHS trajectory, factors associated with increased risk for belonging to SS trajectory were male gender, first child, maternal age and working status, night-waking, parental presence when falling asleep, television-viewing duration and both the "Processed and fast foods" and the "Baby food" dietary patterns at age 2 years. Factors positively associated with the CS trajectory were maternal smoking, bottle-feeding at night and the "Processed and fast foods" dietary pattern at age 2 years whereas child's activity and emotionality scores at age 1 year were negatively associated.

**Conclusion.** We identified distinct NSD trajectories among preschoolers and associated early life factors. Some of them may reflect less healthy lifestyle, providing cues for early multi-behavioral prevention interventions.

**Keywords**: Preschoolers; Group-based trajectory modeling; Sleep duration; Cohort; Epidemiology; Public health;

**HIGHLIGHTS**

- Longitudinal data were analyzed using a data-driven developmental approach
- Five different night sleep duration trajectories were identified in preschoolers
- Specific early life factors were associated with each trajectory
- Most of them were living habits and may reflect global less healthy lifestyle
- Early multi-behavioral prevention interventions may be beneficial

**Abbreviations:**

NSD: night-sleep duration

BMI: body-mass index

SS: short-sleep

MLS: medium-low sleep

MHS: medium-high sleep

CS: changing sleep

LS: long sleep

# 1. INTRODUCTION

Sleep is of vital importance for children's health and wellbeing. There is now accumulating evidence that insufficient quantity and/or quality of sleep have a negative impact on children's physical and mental health development, cognitive function, behavior and academic success [1–4]. Sleep disorders and short sleep duration in childhood have also been suggested as predictors of sleep disorders and short sleep duration in adolescence and adulthood [5,6]. Investigating early determinants of sleep durations may help to better understand physiopathology and develop prevention interventions of unhealthy sleep patterns.

A British longitudinal cohort study interested in factors associated with normal sleep duration variation among over 11,000 children aged 6 months to 11 years, showed that girls consistently slept longer than boys, and that older mother age (>35 years) was associated with shorter sleep duration [7]. Studies focusing on short sleep, with age-specific cut-offs, also showed that girls were less likely to be short sleepers [8–10]. Other factors associated with short sleep were identified as lower socio-economic status, non Caucasian ethnic group, maternal stress and/or depression, prematurity, low birth weight, care outside the home, TV/screen viewing especially before going to sleep and late bedtime [7,8,10,11]. Parental behavior at bedtime (e.g. parental presence until sleep onset, feeding), especially among toddlers, is an additional important risk factor for fragmented sleep and consequently shorter sleep duration [12–15].

A decrease in children's total sleep duration has been reported in the last decades [16,7], suggesting that a growing number of children may now get shorter sleep duration than needed. The American Academy of Sleep Medicine recently recommended a total mean sleep duration of 10 to 13 hours per 24 hours among preschoolers [17]. As reported by Galland et al and Blair et al, standard deviation of the means varies between 1 to 2 hours among preschoolers [7,16] and sleep duration mean may not correctly reflect the variety of sleep durations during childhood. Longitudinal sleep patterns or trajectory may also be of interest. Previous research concerning Canadian preschoolers identified four sleep trajectories between 2.5 and 5 years: short persistent (<9hrs/night), short increasing (<9h up to 3.5 years old and then around 10.5hrs), 10-hrs persistent and 11-hrs persistent patterns [18,19]. The authors showed an increased risk of externalizing problems [18] and high hyperactivity scores [19] in 6 year-old children with short sleep duration trajectory (i.e. short-persistent vs. 11hrs-persistent sleepers). They also reported that risk factors associated with both short sleep trajectory and high hyperactivity scores were male gender, low household income, low maternal education and parental presence during night awakening [19]. However, they did not search for factors associated with the short sleep trajectory by itself or with other sleep trajectories. In addition, European preschoolers tend to sleep longer on average than the North American preschoolers [14,20] and may not display the same sleep patterns over time or associated factors.

This first longitudinal study on children' sleep duration in a French birth-cohort aimed at i) identifying sleep duration trajectories between 2 and 5-6 years; and ii) identifying factors associated with each sleep duration trajectory.

## 2. MATERIALS/SUBJECTS AND METHODS
### 2.1. STUDY DESIGN

The EDEN study aims at investigating the pre- and post-natal determinants of child health and development. Details of the EDEN study protocol have been previously published [21]. Briefly, pregnant women under 24 weeks of amenorrhea were recruited between 2003 and 2006, in the university hospitals of Poitiers and Nancy. Those under 18 years, unable to give informed consent, functionally illiterate in French, with a history of diabetes, planning on changing address or without social security coverage were excluded from the cohort. Multiple pregnancies were also excluded. Among the 3758 women invited to participate, 2002 (53%) agreed to enroll into the study. Due to miscarriages, stillbirths and attrition, 1899 children were enlisted at birth. Written informed consent was obtained twice from parents: at enrolment and after the child's birth. The study was approved by the ethics research committee of Bicêtre Hospital (Comité Consultatif de Protection des Personnes dans la Recherche Biomédicale) and by the Data Protection Authority (Commission Nationale de l'Informatique et des Libertés).

### 2.2. DATA COLLECTION

Data were collected using parental self-administered questionnaires and during clinical examinations, including anthropometric measurements of each child. Sleep items were study-designed ones.

#### 2.2.1. Main measure: Night sleep duration

Night sleep durations were collected at ages 2, 3 and 5-6 years old and were calculated based on the answers to the following questions: "Usually, at what time does your child go to bed?", "Usually, at what time does your child wake up?". Responses were recorded in hours and minutes.

#### 2.2.2. Predictors

The household socio-economic and demographic factors, as well as maternal characteristics were collected at inclusion. Household income was divided into three categories: below €1500 per month (≈ French threshold of poverty for a family), between €1500 and €3000 per month, and above €3000 per month (≈10$^{th}$ upper percentile of French income distribution). Education level was also defined in three categories, using the highest level reached by one of the parents: below high-school diploma, high-school diploma, and above. Single parenting was defined as a mother living without the child's father, another companion, or another adult family member. Mothers reported information on age at delivery and tobacco consumption during and after pregnancy (coded as never, only after pregnancy and always (during ± after pregnancy)). The mother's depressive symptoms during pregnancy were assessed by the French version of the Center of Epidemiologic Studies Depression Scale (CES-D). Mothers with a CES-D score of ≥23 were considered to present depressive symptoms [22]. Body mass index (BMI) before pregnancy was calculated using reported height and weight. The maternity ward of recruitment (Nancy/Poitiers) and the mother's working status at age 2 ~~(coded as yes/no)~~ were also taken into account.

The child's characteristics and anthropometrics were collected at birth from self-reported questionnaires and medical records, including gender, first child (yes/no), ponderal index (defined by birth weight in kg divided by the cube of birth length in meters), and preterm birth (< 37 weeks of amenorrhea). Breastfeeding duration was collected in months from prospective self-administered questionnaires. Temperamental traits, namely activity, shyness, emotionality and sociality were assessed at age 1 using the Emotionality Activity and Sociability scale (EAS) [23].

At age 2, data were collected for several sleep characteristics. Nap duration was assessed through two questions: "Does your child regularly take a nap?" "If yes, what is the mean duration of a nap?". Responses were recorded in hours and minutes. Children who did not nap were coded as 0 hours 0 minutes. Frequent night awakenings at the age of 2 years were defined as waking every other night or more (yes/no) over the week preceding the self-questionnaire completion, in the absence of acute illness. Parental presence when falling asleep was also collected through questions on sleeping habits (place (e.g. living room), parental interaction (e.g. holding hands) and bed sharing). A child was considered to sleep without parental presence when parents reported that he/she felt asleep in his/her own bed without any adult interaction.

At age 2, we collected the number of hours per day spent in physical activity (walking, playing outside) and watching television or other screens during a usual week separately for weekdays, Wednesdays (weekday without preschool in France) and for weekend days. As expected, the mean number of hours per day spent in physical activity was statistically different according to the season when the self-questionnaire was completed. We therefore split this variable into quartiles according to each season at which the questionnaire was completed. As this was not the case for the number of hours spent watching TV, the latter was analysed continuously. We also assessed care arrangement (in large collective settings like preschool or day care centres vs. home care). Feeding at night (bottle- or breast-feeding, yes/no) was collected at 2 and 3 years old. The child's BMI-z-score was calculated using the WHO age and gender standards, and obesity was defined as a BMI-z-score ≥2 SD [24]. We also accounted for dietary patterns previously identified in the EDEN cohort [25]. Briefly, children's dietary intake was collected using a short food frequency questionnaire and three dietary patterns - accounting for 26.8% of the explained variance - were identified at 2 years of age using principal component analysis (PCA). The first pattern, labeled ''Processed and fast foods'', was positively correlated with intake of French fries, processed meat, carbonated soft drinks, chocolate, chips, cookies, pizza, fruit juice, meat, dairy desserts, and ice cream (by descending order of PCA loadings). The second pattern, labeled ''Guidelines,'' was characterized mainly by high consumption frequency of cooked vegetables, rice, fresh fruit, raw vegetables, low-fat fish, potatoes, ham, stewed fruit, and meat. The third pattern, labeled ''Baby foods'', had high positive loadings for baby foods, breakfast cereals, and stewed fruit and negative loadings for raw vegetables and fresh fruit.

## 2.3. STATISTICAL METHODS

### 2.3.1. Group-based trajectory modeling

The group-based trajectory modeling method (PROC TRAJ procedure) developed by Nagin et al. [26] was used to identify meaningful and distinct patterns of sleep duration from the age 2 to 5-6 years. The method is based on the underlying hypothesis that within a population there are inherent groups that evolve according to different sleep patterns. The groups are not directly identifiable or pre-established by sets of characteristics but statistically determined through each series of responses using maximum likelihood. The relationship between age and night sleep duration was modeled by polynomial equations defining trajectories. The most adequate model, regarding the number of groups and the shape of the trajectories, was determined by iterations: different models with two to five groups were computed and then compared using the Bayesian Information Criteria (BIC) and favoring parsimony. The chosen model quality was verified according to the recommended criteria: the average posterior probabilities for each subgroup ($\geq 0.7$), the odds of correct classification ($\geq 5$), and the similarity between the model's estimation of the trajectory prevalence and the actual prevalence [26].

Children were included in the trajectory's elaboration if their parents had answered the questions regarding night-sleep durations at least at two time points out of three. To verify the robustness of the model, sensitivity analysis was performed in children who had complete sleep duration data at all three time points.

### 2.3.2. Study of associated factors

Children were assigned to the trajectory for which he/she has the highest probability of belonging. Multinomial logistic regressions were then computed to study factors associated with modeled night-sleep duration trajectories. We chose as reference trajectory, the one that was the nearest of the recommended sleep duration between 2 and 5 years old, i.e. between 11 and 11.30 hours/night. An unadjusted analysis was first performed on collected data described in the previous section, then a multivariable analysis that included factors with a p-value <0.20 in the unadjusted analyses and socioeconomic factors (i.e. parental education status, household income and recruitment center). Missing values for explanatory factors represented 4.0% of the total data set. Simple imputations (modal value for categorical variables and median value for continuous ones) were implemented.

All analyses were performed using SAS (SAS 9.3 SAS Institute Inc, Cary, NC, USA).

## 3. RESULTS

### 3.1. Night-sleep duration trajectories

Out of the 1899 children enrolled at birth, self-administered questionnaires were available for 1349 children at age 2, 1377 at age 3 and 1255 at age 3. A total of 1205 presented two out of three completed time points for night-sleep duration and were included in the trajectory conception. Compared to included children, non-included children (N=694) were more likely to be born to a mother from Nancy recruitment center (41.4% versus 31.5%, $p<10^{-4}$), with low incomes (19.0% ≤1500 €/months versus 10.4%, p=0.01), low educational level (31.0% versus 15.2% with a level below high-school diploma, $p<10^{-4}$), who smoked during and after pregnancy (66.7% versus 22.3%, $p<10^{-4}$), and who was unemployed 2 years post-partum (43.8% versus 26.0%, $p<10^{-4}$). There was no difference between the two population groups regarding maternal age at delivery (p=0.09), parity (p=0.92), child gender (p=0.46), prematurity status (p=0.82), or breastfeeding duration (p=0.45). Mean night-sleep duration of selected children was 11hrs06 (SD 0h49 - range 8hrs00-14hrs00) at age 2 (N=1082), 10hrs52 (SD 0hrs40 - range 9hrs00-13hrs45) at age 3 (N=1170) and 10hrs52 (SD 0hrs28 - range 9hrs17-12hrs17) at age 5-6 (N=1020).

The optimal and parsimonious trajectory model to describe night-sleep duration patterns was a five-group model as illustrated in Figure 1: a Short-Sleep duration trajectory (SS, always <10hrs30/night) best explained by a quadratic relationship with time and representing 4.9% of the children; a Medium-Low-Sleep duration trajectory (MLS, 10hrs30-11hrs00/night) best explained by a positive linear relationship with time and representing 47.8% of the children; a Medium-High-Sleep duration trajectory (MHS, around 11hrs30/night) best explained by a negative linear relationship with time and representing 37.2% of the children; a Long-Sleep duration trajectory (LS, ≥11hrs30/night) best explained by a negative linear relationship with time and representing 4.5% of the children; and a Changing-Sleep duration trajectory (CS, i.e. up to age 3 similar to LS and then to MLS) estimated to be quadratic over time and representing 5.6% of the children. Nagin's recommended criteria for goodness of fit were met for all groups [26]. To test robustness of the model, the same procedure was performed including children with three completed time points (N=862) that showed no notable difference regarding i) the number of groups, ii) the shape of the trajectories or iii) compliance with recommended criteria for goodness of fit (data not shown). Hence, we chose to include the largest sample size for further analysis.

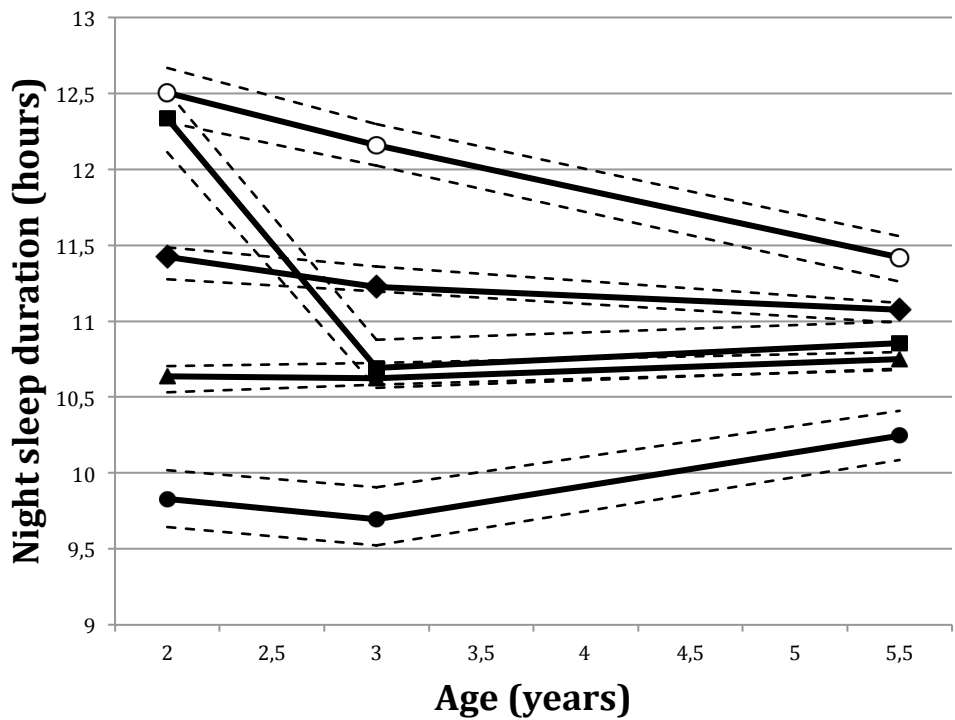

**Figure 1.** Night sleep duration trajectories obtained among the EDEN preschool children (N=1205). Full lines represent mean sleep duration trajectories. Black circles = Short sleepers (SS, 4.9% of the children): triangles = Medium Low sleepers (MLS, 47.8% of the children); diamonds = Medium High sleepers (MHS, 37.2% of the children), squares = Changing (CS, 5.6% of the children) and white circles = Long sleepers (LS, 4.5% of the children). Dashed lines represent the 95% confidence intervals of the trajectory estimations.

### 3.2. Factors associated with night-sleep duration trajectories

Population characteristics are presented in Table 1 and multivariable analysis is reported in Table 2. As compared to the MHS trajectory, membership to the SS trajectory between 2 to 5-6 years was more likely in boys, first-borns, born to older mothers, with at age 2 years higher occurrence of frequent night waking, more frequent parental presence when falling asleep, longer time spent in front of TV, higher scores on both the "Processed and fast food" and "Baby foods" dietary patterns, and when mothers worked 2 years post-partum. Of note, most of these factors were also associated with the MLS trajectory membership, as compared to the MHS. There was no association between membership to SS or to MLS trajectories and the maternal smoking status, the child's temperament scores at age 1 feeding at night at age 2.

**Table 1**. Characteristics of the 1205 children in the full sample and according to each night-sleep duration trajectory. Data are %(n) or mean±sd.

|  | Full sample (n=1205) | SS[a] n=59 | MLS[a] n=576 | MHS[a] n=448 | CS[a] n=68 | LS[a] n=54 | Global p-value |
|---|---|---|---|---|---|---|---|
| **Socieconomic factors** | | | | | | | |
| Center (Nancy) | 46.8 (564) | 47.5 (28) | 46.9 (270) | 45.1 (202) | 60.3 (41) | 42.6 (23) | |
| Parental education level | | | | | | | |
|     < High school | 15.2 (183) | 17.0 (10) | 14.6 (84) | 15.2 (68) | 17.7 (12) | 16.7 (9) | |
|     High school | 18.2 (219) | 25.4 (15) | 17.5 (101) | 17.4 (78) | 22.0 (15) | 18.5 (10) | |
|     > high school | 66.6 (803) | 57.6 (34) | 67.9 (391) | 67.4 (302) | 60.3 (41) | 64.8 (35) | |
| Household income | | | | | | | |
|     <1500 €/month | 10.4 (125) | 8.5 (5) | 9.0 (52) | 10.5 (47) | 19.1 (13) | 14.8 (8) | |
|     1501-3000 €/month | 58.9 (710) | 64.4 (38) | 61.3 (353) | 56.7 (254) | 54.4 (37) | 51.9 (28) | |
|     >3000 €/month | 30.7 (370) | 27.1 (16) | 29.7 (171) | 32.8 (147) | 26.5 (18) | 33.3 (18) | |
| **Maternal characteristics** | | | | | | | |
| Age at delivery (years) | 30.0±4.6 | 30.5±4.59 | 30.5±4.5 | 29.7±4.7 | 28.6±5.0 | 29.4±4.4 | ** |
| Smoking habits | | | | | | | * |
|     Never | 64.3 (775) | 57.6 (34) | 66.8 (385) | 64.7 (290) | 50.0 (34) | 59.3 (32) | |
|     Only after pregnancy | 13.4 (161) | 11.9 (7) | 12.9 (74) | 14.7 (66) | 11.8 (8) | 11.1 (6) | |
|     Always | 22.3 (269) | 30.5 (18) | 20.3 (117) | 20.5 (92) | 38.2 (26) | 29.6 (16) | |
| Pre-pregnancy BMI | 23.1±4.3 | 23. 6±4.4 | 23.3±4.4 | 22.9±4.1 | 23.1±4.8 | 23.0±4.5 | |
| Depressive symptoms[b] | 5.2 (62) | 3.4 (2) | 4.3 (25) | 5.1 (23) | 11.8 (8) | 7.4 (4) | |
| Single parenting | 3.5 (40) | 1.8 (1) | 4.1 (22) | 3.3 (14) | 3.0 (2) | 1.9 (1) | |
| **Child characteristics** | | | | | | | |
| First child | 46.6 (561) | 61.0 (36) | 43.6 (251) | 46.2 (207) | 58.9 (40) | 50.0 (27) | * |
| Gender (Boy) | 53.2 (641) | 71.2 (42) | 56.3 (324) | 50.5 (226) | 36.8 (25) | 44.4 (24) | *** |
| Pre-term birth[c] | 5.3 (64) | 6.8 (4) | 4.9 (28) | 6.0 (27) | 0.0 (0) | 9.3 (5) | |
| Child ponderal index (kg/m$^3$) | 27.0±2.7 | 27.0±2.3 | 26.9±2.8 | 26.9±2.54 | 27.4±3.1 | 27.2±3.2 | |
| Breastfeeding duration (months) | 3.4±3.7 | 2.9±3.9 | 3.5±3.8 | 3.1±3.6 | 3.3±3.9 | 3. 7±4.0 | |
| Temperament at age 1[d] | | | | | | | |
|     Activity | 3.53±0.48 | 3.63±0.44 | 3.53±0.47 | 3.54±0.49 | 3.41±0.53 | 3.51±0.50 | |
|     Shyness | 2.08±0.56) | 1.97±0.42 | 2.08±0.58 | 2.10±0.56 | 2.11±0.62 | 2.12±0.55 | |
|     Emotionality | 2.76±0.70) | 2.8±0.86 | 2.76±0.67 | 2.77±0.69 | 2.63±0.72 | 2.80±0.80 | |
|     Sociability | 3.69±0.59 | 3.83 ± 0.69 | 3.72±0.58 | 3.65±0.58 | 3.62±0.61 | 3.74±0.59 | |
| **Characteristics at 2 years of age** | | | | | | | |
| Working mother | | | | | | | *** |
|     No | 26.0 (313) | 20.3 (12) | 22.7 (131) | 29.2 (131) | 29.4 (20) | 35.2 (19) | |
|     Part-time | 34.1 (411) | 33.9 (20) | 35.6 (205) | 34.2 (153) | 32.4 (22) | 20.4 (11) | |
|     Full-time | 39.9 (481) | 45.8 (27) | 41.7 (240) | 36.6 (164) | 38.2 (164) | 44.4 (24) | |
| Collective care arrangement | 21.1 (254) | 18.6 (11) | 22.6 (130) | 20.8 (93) | 16.2 (11) | 16.7 (9) | |
| Nap duration (hrs/day) | 2hrs04±0hrs30 | 2hrs07±0hrs34 | 2hrs04±0hrs30 | 2hrs05±0hrs29 | 2hrs02±0hrs32 | 2hrs02±0hrs30 | |
| Frequent night-wakings | 20.3 (245) | 40.7 (24) | 24.1 (139) | 14.1 (63) | 19.1 (13) | 11.1 (6) | *** |
| Falling asleep with parental presence | 11.5 (139) | 25.4 (15) | 13.7 (79) | 8.7 (39) | 5.9 (4) | 3.7 (2) | *** |
| Feeding at night | 26.4 (318) | 25.4 (15) | 26.9 (155) | 24.8 (111) | 45.6 (31) | 11.1 (6) | *** |
| Body mass index (z-score) | 0.25±0.98 | 0.28±0.78 | 0.25±0.89 | 0.26±0.85 | 0.27±0.80 | 0.03±0.74 | |
| Television watching (hrs/day) | 0hrs40±0hrs43 | 1hrs10±1hrs10 | 0hrs41 (0hrs40) | 0hrs37 (0hrs39) | 0hrs31 (0hrs40) | 0hrs40 (0hrs49) | *** |
| Physical activity (quartiles)[e] | | | | | | | |
| Q1 | 23.4 (282) | 28.8 (17) | 24.5 (141) | 21.4 (96) | 27.9 (19) | 16.7 (9) | |
| Q2 | 27.2 (328) | 17.0 (10) | 25.4 (146) | 31.2 (140) | 20.6 (14) | 33.3 (18) | |
| Q3 | 24.2 (292) | 27.1 (16) | 24.1 (139) | 23.7 (106) | 23.6 (16) | 27.8 (15) | |
| Q4 | 25.2 (303) | 27.1 (16) | 26.0 (150) | 23.7 (106) | 27.9 (19) | 22.2 (12) | |
| Dietary pattern score | | | | | | | |
| Processed and fast foods | -0.05±0.94 | 0.12±1.08 | -0.01±0.96 | -0.12±0.87 | 0.11±1.07 | -0.22±0.80 | |
| Guidelines | 0.03±0.93 | 0.01±1.05 | 0.01±0.91 | 0.07±0.94 | -0.05±0.9 | 0.11±0.94 | |
| Baby foods | -0.02±0.97 | 0.34±0.94 | -0.03±0.98 | -0.07±0.94 | 0.11±0.93 | 0.03±1.08 | * |

*≤0.05, **≤0.01, ***≤0.001

[a] Night sleep duration trajectories: SS for Short sleep duration, MLS for Medium Low sleep duration, MHS for Medium High sleep duration, CS for Changing sleep duration and LS for Long sleep duration trajectory.

[b] Center of Epidemiologic Studies Depression scale score ≥23

[c] Birth before 37 weeks of amenorrhea

[d] Emotionality Activity and Sociability scale (EAS); score range 0–5, a higher score indicates more activity, shyness, emotionality or sociability

[e] Quartiles of physical activity according to each season

Table 2. Multiple multinomial logistic regression for sleep trajectories, N=1205. The MHS[a] trajectory served as reference

| | SS[a] OR[c] (95% CI) | MLS[a] OR[c] (95% CI) | CS[a] OR[c] (95% CI) | LS[a] OR[c] (95% CI) | Global p-value[b] |
|---|---|---|---|---|---|
| **Socieconomic factors** | | | | | |
| Center (Nancy) | 1.44 (0.77 – 2.70) | 1.19 (0.91 - 1.56) | 2.55 (1.41 - 4.58)** | 0.96 (0.52 - 1.79) | * |
| Parental education level | | | | | |
| < High school | 0.80 (0.30 – 2.14) | 0.91 (0.57 - 1.44) | 0.72 (0.29 - 1.79) | 1.12 (0.41 – 3.06) | |
| High school | (reference) | (reference) | (reference) | (reference) | |
| > high school | 0.58 (0.27 – 1.26) | 1.00 (0.69 - 1.44) | 0.78 (0.37 - 1.66) | 0.94 (0.41 - 2.18) | |
| Household income (euros/months) | | | | | |
| <1500 | 0.67 (0.18 - 2.45) | 1.21 (0.69 - 2.09) | 1.75 (0.63 - 4.89) | 1.20 (0.39 - 3.65) | |
| 1501-3000 | 1.30 (0.62 - 2.74) | 1.38 (1.01 - 1.88) | 1.23 (0.62 - 2.43) | 0.84 (0.41 - 1.71) | |
| >3000 | (reference) | (reference) | (reference) | (reference) | |
| **Maternal characteristics** | | | | | |
| Age at delivery (years) | 1.14 (1.06 - 1.22)*** | 1.06 (1.03 - 1.10)*** | 0.98 (0.91 - 1.05) | 1.00 (0.93 - 1.07) | *** |
| Smoking habits | | | | | |
| Never | (reference) | (reference) | (reference) | (reference) | |
| Only after pregnancy | 1.10 (0.44 - 2.76) | 0.89 (0.61 - 1.30) | 0.91 (0.39 - 2.13) | 0.84 (0.33 - 2.13) | |
| Always | 1.92 (0.95 - 3.87) | 1.03 (0.73 - 1.44) | 2.40 (1.27 - 4.55)** | 1.60 (0.79 - 3.23) | |
| Depressive symptoms[d] | 0.75 (0.16 - 3.44) | 0.85 (0.47 - 1.55) | 2.63 (1.03 - 6.73)* | 1.33 (0.43 - 4.14) | |
| **Child characteristics** | | | | | |
| First child | 3.19 (1.57 - 6.50)*** | 1.08 (0.80 - 1.45) | 1.83 (0.96 - 3.49) | 1.00 (0.51 – 1.96) | ** |
| Gender (Boy) | 2.46 (1.30 - 4.68)** | 1.28 (0.99 - 1.66) | 0.60 (0.34 - 1.06) | 0.84 (0.45 - 1.50) | ** |
| Temperament at age 1[e] | | | | | |
| Activity | 1.03 (0.52 - 2.01) | 0.91 (0.69 - 1.21) | 0.52 (0.30 - 0.91)* | 0.83 (0.45 - 1.55) | |
| Shyness | 0.91 (0.49 - 1.70) | 1.02 (0.79 - 1.33) | 1.12 (0.66 - 1.91) | 1.14 (0.63 - 2.06) | |
| Emotionality | 0.85 (0.54 - 1.34) | 0.93 (0.76 - 1.13) | 0.62 (0.41 - 0.95)* | 1.00 (0.63 - 1.58) | |
| Sociability | 1.64 (0.93 - 2.87) | 1.19 (0.93 - 1.52) | 1.05 (0.63 - 1.75) | 1.46 (0.84 - 2.52) | |
| **Characteristics at 2 years of age** | | | | | |
| Working mother | | | | | ** |
| No | (reference) | (reference) | (reference) | (reference) | |
| Part-time | 2.39 (0.99 – 5.74) | 1.42 (1.00 – 2.01) | 1.18 (0.57 – 2.42) | 0.48 (0.21 – 1.08) | |
| Full time | 2.93 (1.26 – 6.80)* | 1.78 (1.26 – 2.52)** | 1.22 (0.59 – 2.50) | 0.98 (0.49 – 1.96) | |
| Frequent night-wakings | 3.71 (1.90 - 7.21)*** | 1.91 (1.35 - 2.70)** | 1.47 (0.72 – 2.99) | 0.77 (0.31 - 1.93) | *** |
| Falling asleep with parental presence | 3.44 (1.59 - 7.41)** | 1.62 (1.05 - 2.50)* | 0.54 (0.17 - 1.72) | 0.50 (0.11 – 2.22) | ** |
| Feeding at night | 0.55 (0.27 - 1.14) | 1.00 (0.73 - 1.35) | 2.46 (1.38 - 4.47)** | 0.37 (0.15 - 0.92)* | *** |
| Television watching (hrs/day) | 2.11 (1.50 - 2.97)** | 1.13 (0.93 - 1.38) | 0.57 (0.35 - 0.93)* | 1.13 (0.73 - 1.73) | *** |
| Dietary pattern score | | | | | |
| Processed and fast foods | 1.45 (1.06 – 2.01)* | 1.21 (1.04 - 1.41)* | 1.38 (1.04 - 1.81)* | 0.83 (0.57 - 1.22) | * |
| Guidelines | 1.00 (0.73 - 1.38) | 0.94 (0.82 - 1.09) | 0.91 (0.68 - 1.22) | 1.07 (0.78 - 1.47) | |
| Baby foods | 1.48 (1.08 - 2.03)** | 1.06 (0.93 - 1.22) | 1.18 (0.89 - 1.57) | 1.12 (0.83 - 1.51) | |

*≤0.05, **≤0.01, ***≤0.001

[a] Night sleep duration trajectories: SS for Short sleep duration, MLS for Medium Low sleep duration, CS for Changing sleep duration and LS for Long sleep duration trajectory.

[b] p-value for the global effect of the corresponding factor analyzed

[c] odds ratio (OR) and 95% confidence interval (95%CI)

[d] Center of Epidemiologic Studies Depression scale score ≥23

[e] Emotionality Activity and Sociability scale (EAS); score range 0–5, a higher score indicates more activity, shyness, emotionality or sociability

As compared to the MHS trajectory, membership to the CS trajectory was more likely in children from Nancy, fed at night, with higher scores on the "Processed and fast foods" dietary pattern at 2 years, as well as and in children whose mother presented depressive symptoms during pregnancy and smoked during and after pregnancy. The activity and emotionality temperament score at 1 year old decreased the likelihood of membership to the CS trajectory as did time spent in front of TV. Moreover, additional adjustment for night feeding at 3 years did not change the results. In particular the OR for night feeding stayed very stable at 2 years (OR=2.28 (95%CI 1.19-4.40)), the one for night feeding at 3 years was borderline significant (OR=1.93 (0.95-3.88))

Feeding at night at 2 years old was the only factor associated with membership to the LS trajectory showing that night fed children were less likely to belong to this trajectory than to the MHS one.

No association was observed between sleep duration trajectories and socioeconomic factors (education level, household incomes).

## 4. DISCUSSION

This study, using trajectory modeling, gives new insight into developmental patterns of night-sleep duration and associated factors among preschoolers. In this context, three trajectories and their associated factors are of particular interest: the SS trajectory that was under the sleep duration recommendations, the LS trajectory that slowly decreased between 12.5hrs/night to 11hrs/night between 2 and 5-6 years old and the CS trajectory that showed a 2h rapid decrease between 2 and 3 years.

### 4.1. Sleep duration trajectories

We identified five night sleep duration trajectories among preschoolers. We showed, as others, that each night sleep duration trajectory (except the CS one) was quite stable during early childhood [18,27]. This may reflect one child's inherent sleep needs, however we identified several factors associated with the belonging to each trajectory (discussed below) stressing the importance for good sleep habits setting in early infancy. In contrast, the thresholds allowing to distinguish different trajectories were overall higher than those described in Canadian pre-school population (short was defined as <9hrs/night, medium low as 10-hrs persistent, medium high as 11-hrs persistent, changing as short then between 10-hrs and 11-hrs persistent, and long were not observed) [18,19]. These differences may partly be explained by the overall higher night- and total-sleep durations in French [14] and in North-European [20] preschool children, as compared to North Americans [18,19,27]. In the present study, mean sleep durations at age 2, 3 and 5-6 were in the higher limits of the American Academy of Sleep Medicine recommendations to promote optimal health [17]. However, we reported similar proportion of members of the "short" sleep trajectory group (5 to 6%) to that published among children of the same ages using the same trajectory method [18,19].

### 4.2. Factors associated with short sleep duration trajectory

The present study confirmed some of the risk factors for short sleep duration suggested in both cross-sectional and longitudinal studies among preschool or school aged children, i.e. male gender [7–10,14,19] and maternal older age [7,14]. First-born children have been shown to present shorter sleep duration in infancy [28], which could suggest higher parental stress or higher parental intervention on child's sleep or both. However, the relation was observed in the present analysis independently of maternal age, maternal depressive status and sleep habits. A couple of sleep habits, already shown to be associated with short sleep durations [27], were positively associated with SS trajectory, namely parental presence when falling asleep and frequent night waking. As often discussed, the association between child's sleep and sleep strategies may be bidirectional [29], and the child's temperament may influence sleep and sleep behaviors [30,31]. However, the observed relation persisted after adjustment for child's temperament scores suggesting involvement of other factors such as parental beliefs and practices [32]. Altogether, these elements indicate that children belonging to the short sleep trajectory could benefit from behavioral interventions promoting healthier sleep hygiene, shown to be associated with longer sleep duration and less night-waking in childhood [33,34].

Time spent in front of TV at age 2 was positively associated with this SS trajectory between 2 and 5-6 years old, as observed by others [35]. Beside children with low physiological sleep needs who may be entertained by longer TV watching, this might be explained either by direct replacement of sleep by TV watching or by the program viewed and its aggressiveness associated with more sleep troubles [36]. Unfortunately, these hypotheses could not be addressed in the current study, given that the appropriate information was not collected. Another explanation could be the blue light exposure from the mobile screens altering circadian rhythms. The low access to these devices when data were collected makes this hypothesis unlikely. Children belonging to the SS trajectory more often had working mothers at age 2. The latter may be more likely to come back home later than their non-working counterparts, making it more challenging to get their children to sleep early; they also may need to wake them up earlier for care arrangement, altogether resulting in shorter night sleep duration [37,38]. However, in the current study, there was no difference in prevalence of care arrangement according to the five trajectories. Finally, shorter and lower quality of sleep have been associated with higher energy intake mainly through high fat food in adults [39] and children [40]. A recent review also noted that diet, especially diet rich in fresh fruits, vegetables, whole grains, and low-fat protein sources, promotes sleep quantity in adults [41]. In children, Kocevska et al. showed that a higher contribution of fat to total energy intake at 13 months of age was negatively associated with sleep duration at 2 years of age [42]. These findings are in line with the positive association between "Processed and fast foods" dietary pattern scores supposedly positively associated with total energy intake and SS trajectory observed in the current study. A positive association is also noted with the "Baby foods" score. One may say that working mothers with less available time to cook may favor prepared foods for their child either processed and fast foods or more specifically baby food [43–45]. In this context, these latter associations may reflect a less healthy lifestyle. Risk factors associated with the MLS trajectory (between 10hrs30 and 11hrs/night) were similar to those associated with the SS trajectory except for gender, first child, television-viewing duration and baby-food dietary patterns when compared to MLH trajectory. MLS trajectory is, however, within the American Academy of Sleep Medicine recommendation of sleep duration range to promote optimal health among preschoolers. Further studies will be needed to investigate the relationships between these sleep trajectories and subsequent health outcomes.

### 4.3. Factors associated with changing night-sleep duration trajectory

Factors associated with CS trajectory (as compared with the MHS) were quite specific and may reflect factors associated with the sharp decline in sleep duration between 2 and 3 years old. Of note, care-arrangement, nap duration, working status, incomes and parental education levels were not associated with this trajectory. Consistent with other studies [11], we found that maternal depression was associated with shorter sleep duration in children. Several studies conjecture that at least part of this link is mediated by inappropriate sleep strategies [46], e.g. feeding at night (as also observed here). Another inappropriate strategy could be the use of processed and fast foods for the child, that may bring too much fat shown to be deleterious for good sleep in childhood [42]. Processed and fast foods dietary pattern scores have been shown to track (spearman correlation=0.40, p<0.001) between 2 and 3 years old in the EDEN study [25]. We found that easy child temperament at 1 year old (lower score for activity and emotionality) was positively associated with the CS trajectory between 2 and 5-6 years old, independently of maternal depressive status. Maternal smoking, especially during pregnancy, has been associated with sleep disturbances [47,48], through prenatal physiological modification involving hypoxia (as those involved in sudden infant death syndrome [49] or by a direct increase of health problems such as asthma, lower respiratory function, infections (reviewed in Treyster et al [50]) or both. Overall, factors associated with membership to the CS trajectory were behavioral ones reflecting a global life style less stringent with familial health and health recommendations. The positive association between Nancy recruitment center and the CS trajectory is unexpected and still unexplained. It may reflect incomplete accounting for some risk factors (e.g. socioeconomic or environmental factors) or existence of some yet unidentified ones within this specific group of children. Future studies will be needed to explore in particular lifestyle changes that occurred between 2 and 3 years old that might explain, at least partially, the rapid decrease of sleep duration within this sleep trajectory.

### 4.4. Factors associated with long night sleep duration trajectory

After adjustment, the only factor remaining significantly associated to the LS trajectory was a negative one. Children within this trajectory were less ~~bottle~~ fed at night at two years old than the children of the MHS trajectory. This confirms the reduced occurrence of night waking and thus of strategies involved to get the child go back to sleep such as feeding. Of note, there was no association with the parental education level, familial incomes, temperament scores of the child at 1 year, maternal working status or dietary pattern scores. This suggests that, despite night sleep duration evolution between 2 and 5-6 years, which may reflect children physiological needs in general population; children of both MHS and LS trajectories were quite similar.

### 4.5. Strengths and limitations

Strengths of this study are the general population sample and the longitudinal data for children sleep duration, along with the method used that allowed a powerful developmental analysis of night-sleep duration in preschool age. However, the results of the present study should be interpreted in light of some limitations. Sleep duration was calculated based on self-administered questionnaires asking for usual bedtimes and wake up times with no information on sleep onset. They may be different from bedtimes and difficult for parents to estimate. While sleep duration here reflects time in bed, this is a measure still widely used in epidemiology [7,16,20]. To better estimate sleep onset and time spent awake, an independent and objective measure of sleep by actigraphy would have been more accurate but had not been considered for cost and logistical reasons. Simple parental reports on their child's sleep duration have however been shown to be reliable when compared to actigraphy [51,52]. Parental-reported sleep durations are usually overestimated and night awakenings underestimated compared to actigraphy. However, the questionnaires and methods used allow comparison with the international literature and especially studies performed in Canada with similar sleep questions and trajectory modeling [18,19]. We selected children with available sleep data at least for 2 out of 3 time points leading to sleep duration trajectories estimation on 66% of the original cohort. Bias cannot be ruled out if for example, those who were lost to follow up were more likely to have certain sleep duration characteristics. Comparison of included and excluded children showed no differences for night sleep duration when available and we hypothesize a low impact on the modeled night sleep duration trajectories. However, recruitment rate and attrition in the cohort follow up, lead to a studied population with higher socioeconomic status and higher education level than the targeted population [21], generalization of the findings is not possible especially regarding associated factors to low social classes. This population selection may also explain certain observed non-significant associations. In addition, we analyzed simultaneously 5 trajectories and a large number of variables that may have lead to lack of power especially when considering trajectories with small sample size. Lastly, we did not perform correction for multiple testing and may have falsely identified association between studied factors and sleep duration trajectories. Our results, issued from this first longitudinal study on sleep duration trajectories and associated factors among preschoolers, will need to be confirmed

5. **CONCLUSION**

Trajectory analyses give new insight into developmental patterns of night-sleep duration and associated factors among preschoolers. The study confirmed known early life factors associated with shorter sleep duration in preschoolers but also highlighted new ones such as less healthy dietary patterns. Interestingly, some of these risk factors including dietary patterns were associated with medium-low sleep duration trajectory that however meets the American Academy of Sleep Medicine sleep duration recommendations in preschoolers. This study also identified a particular sleep duration trajectory, the "changing" one, that presented specific early risk factors, including depressive symptoms during pregnancy, tobacco smoking during and after pregnancy, feeding at night and processed dietary pattern at 2 years old. Altogether, identified risk factors associated with shorter or changing sleep duration trajectories between 2 to 5-6 years old were mainly living habits and may reflect global less healthy lifestyle. Thus early multi-behavioral prevention interventions may be beneficial in these populations. Further studies are needed to investigate whether those sleep trajectories are differently related to subsequent health outcomes and to replicate the results in larger mother-child cohorts.

**ACKNOWLEDGMENTS**

Collaborators: We thank the EDEN mother-child cohort study group (I. Annesi-Maesano, J.Y Bernard, J. Botton, M.A. Charles, P. Dargent-Molina, B. de Lauzon-Guillain, P. Ducimetière, M. de Agostini, B. Foliguet, A. Forhan, X. Fritel, A. Germa, V. Goua, R. Hankard, B. Heude, M. Kaminski, B. Larroque†, N. Lelong, J. Lepeule, G. Magnin, L. Marchand, C. Nabet, F. Pierre, R. Slama, M.J. Saurel-Cubizolles, M. Schweitzer, O. Thiebaugeorges).

We thank all funding sources for the EDEN study: Foundation for medical research (FRM), National Agency for Research (ANR), National Institute for Research in Public health (IRESP: TGIR cohorte santé 2008 program), French Ministry of Health (DGS), French Ministry of Research, INSERM Bone and Joint Diseases National Research (PRO-A) and Human Nutrition National Research Programs, Paris–Sud University, Nestlé, French National Institute for Population Health Surveillance (InVS), French National Institute for Health Education (INPES), the European Union FP7 programs (FP7/2007-2013, HELIX, ESCAPE, ENRIECO,Medall projects), Diabetes National Research Program (in collaboration with the French Association of Diabetic Patients (AFD), French Agency for Environmental Health Safety (now ANSES), Mutuelle Générale de l'Education Nationale complementary health insurance (MGEN), French national agency for food security, French speaking association for the study of diabetes and metabolism (ALFEDIAM).